\documentclass[preprint,showpacs,preprintnumbers,amsmath,amssymb]{revtex4}
\usepackage{amsmath}
\usepackage[normalem]{ulem}
\usepackage{xcolor}
\usepackage{float}
\usepackage{graphicx}
\usepackage[noabbrev]{cleveref}

\begin{document}
	
\preprint{}
	
	\title{A validity check of the KATIE parton level event generator in the  $k_t$-factorization and collinear frameworks}
\author{\textit{R. Kord Valeshabadi}}
\affiliation {Department of Physics, University of $Tehran$,
1439955961, $Tehran$, Iran.}
\author{\textit{M. Modarres} }
\altaffiliation {Corresponding author, Email:
mmodares@ut.ac.ir,Tel:+98-21-61118645, Fax:+98-21-88004781.}
\author{\textit{S. Rezaie}}
\affiliation {Department of Physics, University of $Tehran$,
1439955961, $Tehran$, Iran.}
 \date{\today}
	\begin{abstract}
In the present paper, we check and study the validity of the \textsc{KaTie} parton level event generator, by calculating the  inclusive electron-proton (ep) dijet and the Proton-proton (p-p)  Drell-Yan electron-pair productions differential cross sections in the $k_t$-  and collinear factorization frameworks.  The Martin-Ryskin-Watt (MRW) unintegrated parton distribution functions  (UPDFs) are used as the input UPDFs.  The results are compared with  those of  ZEUS ep inclusive dijet and   ATLAS p-p Drell-Yan electron-pair productions, experimental data. The \textsc{KaTie} parton level event generator can directly calculate the cross sections in the $k_t$-factorization framework. It is noticed that the lab to the Breit transformation in this generator is not correctly implemented by its author, so the produced output does not cover the ep ZEUS experimental data in which the mentioned transformation is applied. By fixing the  above transformation  in the \textsc{KaTie}   generator code, we  could appropriately produce the ep inclusive dijet differential cross section, in comparison with those of ZEUS data. It is also shown  that the MRW at the NLO level, with the angular ordering constraint can successfully predict the ATLAS p-p Drell-Yan data.  Finally, as it is expected, we conclude that the $k_t$-factorization is an appropriate tool for the small longitudinal parton momenta and high center of mass energies, with respect to the collinear one.
\end{abstract}

	\maketitle
\section{Introduction}
Monte Carlo event generators are essential tools for experimentalists and theoreticians who attempt to simulate hadronic collisions. These generators are mostly based on the assumption of   collinear factorization, in which the partons are assumed to  behave collinearly in the hadrons. However, generally, this assumption is not correct and parton should be allowed to have transverse momenta ($k_t$). Hence the $k_t$-factorization \cite{dok,Collins:2011zzd} comes into play, in which the transverse momenta of partons can be considered as an extra variable  into the hard interaction calculations. This framework allows to calculate the differential cross sections for different processes by using    unintegrated parton distribution functions (UPDFs).  

One of the parton level event generators that can calculate the cross sections for arbitrary final state particles, in the above framework, is the \textsc{KaTie} \cite{KATIE}. The  program  of this generator can produce Les Houches Event File (LHEF)  \cite{LHEF}, where thereafter, this LHEF file can be passed to CASCADE3 \cite{CASCADE3} event generator to perform parton showering models. The \textsc{KaTie} due to its simplicity   has this  capability to put into use for calculating different differential cross sections with the various    UPDFs. So, it is a suitable choice to perform phenomenological studies of these distributions, as well. Therefore, in order to obtain reliable results, one task is to test this generator for different processes and comparing the results to the experimental data. Hence, checking the validity of this parton level event generator at this early stage is crucially important for future studies of the $k_t$-factorization framework. 
 
The $k_t$-factorization framework considers a more realistic picture of partons inside proton with respect to collinear factorization, and hence it is expected to give a better description of the experimental data, especially in high energies (much greater than few TeV) . Although, due to complications which arise when the transverse momentum of parton come into play, obtaining a fully transverse momentum dependent evolution equation is a challenging task. For example the CCFM evolution equation \cite{CCFM1,CCFM2,CCFM3,CCFM4} which is based on the angular ordering of soft gluon emissions is not defined for all quark flavors. However, recently, another approach which is based on the parton branching method \cite{PB1,PB2}, allows one to obtain the UPDFs naturally by solving the DGLAP evolution equations \cite{DokshitzerDeepInelastic,gribovDeepInelastic,altarelli}. With this method one can also obtain the UPDFs for both quarks and gluon and it is shown to have successful predictions \cite{PB_dynamical_res,PB_application}. 

Another approach for defining UPDFs is the Martin-Ryskin-Watt (MRW) approach \cite{MRW,MRW1} which assumes parton evolves according to the DGLAP evolution equations  \cite{DokshitzerDeepInelastic,gribovDeepInelastic,altarelli} till the last parton emission, and then it resums the no-emission parton probability to the hard factorization scale via the Sudakov form factor. This method is investigated in detail \cite{Mod1,Mod2,Mod3,Mod4,Mod5, Mod6,Mod7} and is  shown to be successful   in describing the data of different processes at the LHC, Tevatron etc \cite{Mod8,Mod9,Mod13,Mod14,Mod15,Mod16,Lipatov_photon,Baranov_Drell,lipatov_jet}. 

Recently, the \textsc{KaTie} generator  is extended its application to the electron(positron)-proton (ep) collision \cite{vanHameren:2019wzx}. Therefore, for the first time we calculated the inclusive dijet production differential cross sections  in the $k_t$-factorization framework and compared our results to those of  ZEUS collaboration data \cite{inc-dijet-mod,Zeus_2010}. It was observed that our results  \cite{inc-dijet-mod} cannot describe the data in a satisfying way, and hence we found that the lab to Breit transformation was not appropriately implemented  in this parton level generator.

In the present work, it is  intended to study the prediction of the MRW formalism at the leading (LO) and next-to-leading order (NLO) levels for two different processes, i.e.  the electon(positron)-proton  (ep) inclusive dijet and proton-proton (p-p) Drell-Yan differential cross sections which can be compared with the ZEUS \cite{Zeus_2010} and ATLAS \cite{Aad:2019wmn} collaborations data, respectively. The \textsc{KaTie} parton level event generator is used, in order to check the validity of this generator by correcting  the  Breit transformation   and also investigating  the MRW method.  

So, the paper is organized as follows: In the section II the theoretical framework is presented which includes the \textsc{KaTie} parton level event generator and the MRW formalism. The results and discussions are given in the section III and the section IV is devoted to the conclusions.
\section{The Theoretical Framework}
	In this section, it is intended to first explain the \textsc{KaTie} parton level event generator, and then discuss about the UPDFs which are used to obtain the  differential cross sections for the ep inclusive dijet  and the  p-p  Drell-Yan processes to be compared with those of  the ZEUS and  ATLAS   collaboration data, respectively. 
\subsection{The \textsc{KATIE} Parton Level Event Generator}
This parton level event generator is mainly composed of four parts, i.e, the input file,  the optimization stage, the event generation, and the histograms creation. In the input file of this histogram, one should write information about the sub-processes, the factorization and renormalization scales, the experimental cuts, the off-shellness or   on-shellness of partons, the PDFs, the UPDFs, the order of non-QCD couplings and the energies of incoming particles. Furthermore, it is also possible to calculate the multi-parton scattering, instead of the single one. It is also worth to mention, that one can use desired UPDFs  by providing grid files in columns of $ln(x)$, $ln(k_t^2)$, $ln(\mu^2)$ and $f_a(x,k_t^2,\mu^2)$ for each parton flavor. Additionally, it is  possible to directly use UPDFs grid files of TMDLIB \cite{TMDLIB}. 

After providing an input file, an optimization of all sub-processes and event generation should be performed, which are not of interest for the end-user. Finally, one can obtain differential cross sections by a FORTRAN file with the name "create$\_$eventfile.f90". In this file, the histograms of interests are developed and the program can read the recorded events in a file which is called raw file and create the distributions of interests. Additionally, in this part one can produce LHEF file, which can itself with the help of CASCADE \cite{baranov2021cascade3} to make showering and hadronisation.

Therefore, this generator is a beneficial phenomenological tool to investigate the $k_t$-factorization framework, and different   UPDFs models. In the next subsection, we give an overview of the MRW UPDF models at the LO and NLO levels which will be used through   this report.
	
	\subsection{The MRW UPDFs}
	\label{MRWUPDFs}
	The MRW UPDFs at the leading order level (LO-MRW UPDFs) are developed by Martin, et al \cite{MRW,MRW1}, and are based on the DGLAP evolution equations. In this framework, by choosing the factorization scale of the DGLAP evolution equation to be the transverse momentum of the parton, a parton evolves collinearly in the proton, until it reaches the last evolution step. In this step, the parton which has the transverse momentum $k_t$, emits a parton and evolves to the factorization scale $\mu$ without any further real emissions. Therefore, the LO-MRW UPDFs are defined as follows:
	
	\begin{equation}
	\label{eq:1}
	f_a(x,k_t^2, \mu^2) = T_a(k_t^2, \mu^2) \frac{\alpha_{s}^{LO}(k_t^2)}{2\pi}\sum_{b=q,g} \int_x^1 \Big[ P_{ab}^{LO}(z)f_b^{LO}(\frac{x}{z}, k_t^2)\Big]\;\mathrm{d}z,
	\end{equation}
	where $f_b^{LO}(\frac{x}{z}, k_t^2)$ in the above equation is the momentum weighted parton density at the leading order (LO), and can be either $\frac{x}{z}q^{LO}(\frac{x}{z}, k_t^2)$ or $\frac{x}{z}g^{LO}(\frac{x}{z}, k_t^2)$ for quark (anti-quark) or gluon, respectively. Additionally, $T_a(k_t^2, \mu^2)$ is   the Sudakov form factor and resums no real emissions probability from the scale $k_t^2$ to $\mu^2$ and which is:
	\begin{equation}
\label{eq:2}
T_a(k_t^2 \leq \mu^2,\mu^2) = exp\left(-\int_{k_t^2}^{\mu^2} \frac{\mathrm{d}\kappa_t^2}{\kappa_t^2} \frac{\alpha_{s}^{LO}(\kappa_t^2)}{2 \pi}\sum_{b=q,g}\int_{0}^1 \xi P_{ba}^{LO}(\xi)\;\mathrm{d}\xi \right).
\end{equation}
It should also be noted that the Sudakov form factor is defined for the $k_t^2 \leq \mu^2$ and for the $k_t>\mu$, $T_a\to1$.

One should note that the LO-MRW UPDFs, in the equations (\ref{eq:1}) and (\ref{eq:2}), which can be defined for all quark flavors $f_q(x,k_t^2, \mu^2)$ and gluon, $f_g(x,k_t^2, \mu^2)$,   are divergent at $z \to 1$ and $\xi \to 1$ for the diagonal terms, i.e. $P^{LO}_{qq}(z)$, $P^{LO}_{gg}(z)$, $P_{qq}^{LO}(\xi)$, $P^{LO}_{gg}(\xi)$. To avoid such divergences which happen as a result of soft gluon emissions, one can either use the angular ordering of the soft gluon emissions or the strong ordering of partons, along the evolution ladder, to put a cutoff on $z$ and $\xi$. The authors of this model, adopted the angular ordering of the soft gluon emissions, hence we use the same cutoff here, i.e,:
  	\begin{equation}
	\label{eq:3}
	z_{max} = \dfrac{\mu}{(k_t+\mu)}, \;\;\;\;\;\; \xi_{max} = \dfrac{\mu}{(\kappa_t+\mu)}.
	\end{equation}
	The $z_{max}$ and $\xi_{max}$  are the maximum allowed values of z and $\xi$. Therefore one should put a Heaviside step functions for the diagonal terms of the equations (\ref{eq:1}) and (\ref{eq:2}), i.e, $\Theta(z_{Max}-z)$ and  $\Theta(\xi_{Max}-\xi)$, respectively. 

It should be noted that the LO-MRW formalism is limited to the $k_t \ge \mu_0$, and hence for defining such distribution in the limit  $k_t < \mu_0$, the density of the partons are assumed to be constant at fixed $x$ and $\mu^2$ in the reference \cite{MRW,MRW1}, and satisfy the normalization condition. Therefore one obtains the density of partons in the limit $k_t < \mu_0$, as:
\begin{equation} \label{eq:seven}
\frac{1}{k_t^2} f_a(x, k_t^2 <\mu_0^2, \mu^2) = \frac{1}{\mu_0^2}a(x,\mu_0^2)T_a(\mu_0^2,\mu^2),
\end{equation}
where in the above equation we fix $\mu_0 = 1 \; GeV$.

The LO-MRW formalism is also extended to the next-to-leading order level, (NLO-MRW), in which instead of the scale $k_t^2$, the virtuality of the parton along the evolution ladder is used, i.e. $k^2=\dfrac{k_t^2}{(1-z)}$. Additionally, the PDFs, the strong ordering coupling constant, and the splitting functions are at the NLO level. However, it is shown in \cite{MRW}  with the use of NLO strong ordering coupling constant along with the NLO-PDFs one can obtain results close to those use in the fully NLO case. For   simplicity the first form is used, where the PDFs and strong ordering coupling constant are at the NLO level, but the splitting functions are at the LO level. Therefore one can write the NLO-MRW UPDFs as:
\begin{equation}
	\label{eq:4}
	f_a(x,k_t^2, \mu^2) =  \sum_{b=q,g} \int_x^1 \frac{\alpha_{s}^{NLO}(k^2)}{2\pi} T_a(k^2, \mu^2) \Big[ P_{ab}^{LO}(z)f_b^{NLO}(\frac{x}{z}, k^2)\Big] \Theta(\mu^2-k^2) \;\mathrm{d}z,
	\end{equation}
where, in the above equation, we implicitly insert, once again, a Heaviside step functions to avoid the soft gluon divergence. There is also an additional cutoff $\Theta(\mu^2-k^2)$ in the above equation,   which limits the virtuality in the region of $k^2 < \mu^2$. Because, $k^2=\dfrac{k_t^2}{(1-z)}$,  in the limit of large fractional momenta this scale can exceed the factorization scale and such a cutoff can prevent it. Furthermore, this cutoff limits the transverse momentum of parton to the region less than factorization scale, in contrast to the LO case, where the partons can freely have transverse momentum larger than the factorization scale.  Using this cutoff, however has a huge negative effect in the smaller center of mass energies, i.e., when the large fractional momenta can play significant role. Because of this small center of mass energy, $z$ and $k^2$ becomes larger and the strong ordering cutoff $\Theta(\mu^2-k^2)$ can suppress quarks and gluon distributions. It is also obvious that due to the dependency of factorization scale and the Sudakov form factor on $z$, one should move them to the argument of   $z$ integral. Finally, the Sudakov form factor in this model is:
\begin{equation}
	\label{eq:5}
	T_a(k^2,\mu^2) = exp\left(-\int_{k^2}^{\mu^2} \frac{\mathrm{d}\kappa^2}{\kappa^2} \frac{\alpha_{s}^{NLO}(\kappa^2)}{2 \pi}\sum_{b=q,g}\int_{0}^1 \xi P_{ba}^{LO}(\xi)\;\mathrm{d}\xi \right).
	\end{equation}
\section{Results and Discussion}
In this section, we intend to calculate the differential cross sections in the $k_t$-factorization frameworks  with   the LO and  NLO-MRW UPDFs by  utilizing the \textsc{KaTie} parton level event generator. In this calculation, we calculate and provide the grid files of our UPDFs for each quark flavor and gluon, using the MMHT2014lo68cl and MMHT2014nlo68cl PDFs \cite{harland-lang_uncertainties_2015,harland-lang_charm_2016} of LHAPDF6 library \cite{LHAPDF6}, as an input PDFs for our model in the equations (\ref{eq:1}) and (\ref{eq:2}). It should also be noted that   the UPDFs  are divided by $k_t^2$, due to the definition of differential cross section in the \textsc{KaTie} generator.
\subsection{KATIE results for the ZEUS inclusive dijet production data}
The ZEUS collaboration data \cite{Zeus_2010} are collected from the collisions of the protons with energy $920\;GeV$ and electrons or positrons with energy $27.5 \; GeV$. The photon virtuality $Q^2$ is between $125 \; GeV < Q^2 < 20000\; GeV$. The minimum transverse energy of the two jets in the Breit frame should be larger than $E_{T,B} > 8\; GeV$ and the invariant mass of the two jets required to be greater than $M_{jj} > 20\; GeV$. The inelasticity is between $0.2 < y < 0.6$. Additionally, at least two jets are required to have    the pseudo-rapidity in the range $-1 < \eta^{lab} < 2.5$, in the lab frame.
	
We can simply perform such calculation with the help of the \textsc{KaTie}  parton level generator \cite{KATIE}. For calculation of the dijet production, we considered the sub-processes $\gamma^\ast + q \to q + g$ and $\gamma^\ast + g \to q + \overline{q}$ with the LO-MRW and NLO-MRW UPDFs.  We also set the factorization and the renormalization scales $\mu_F=\mu_R=Q$, with $n_f=5$.
	
\begin{figure}
		\includegraphics[width=8cm, height=9cm]{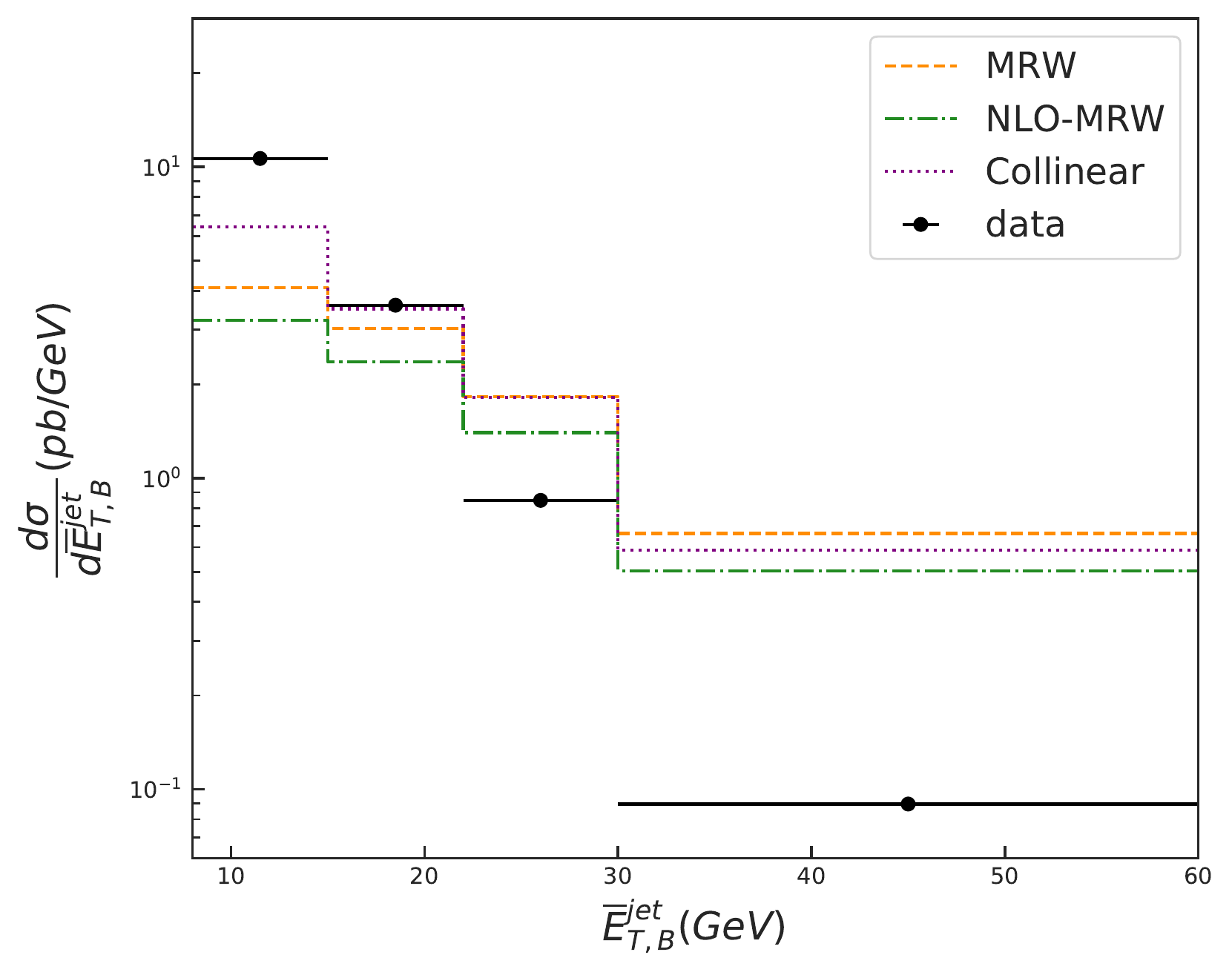}
		\includegraphics[width=8cm, height=9cm]{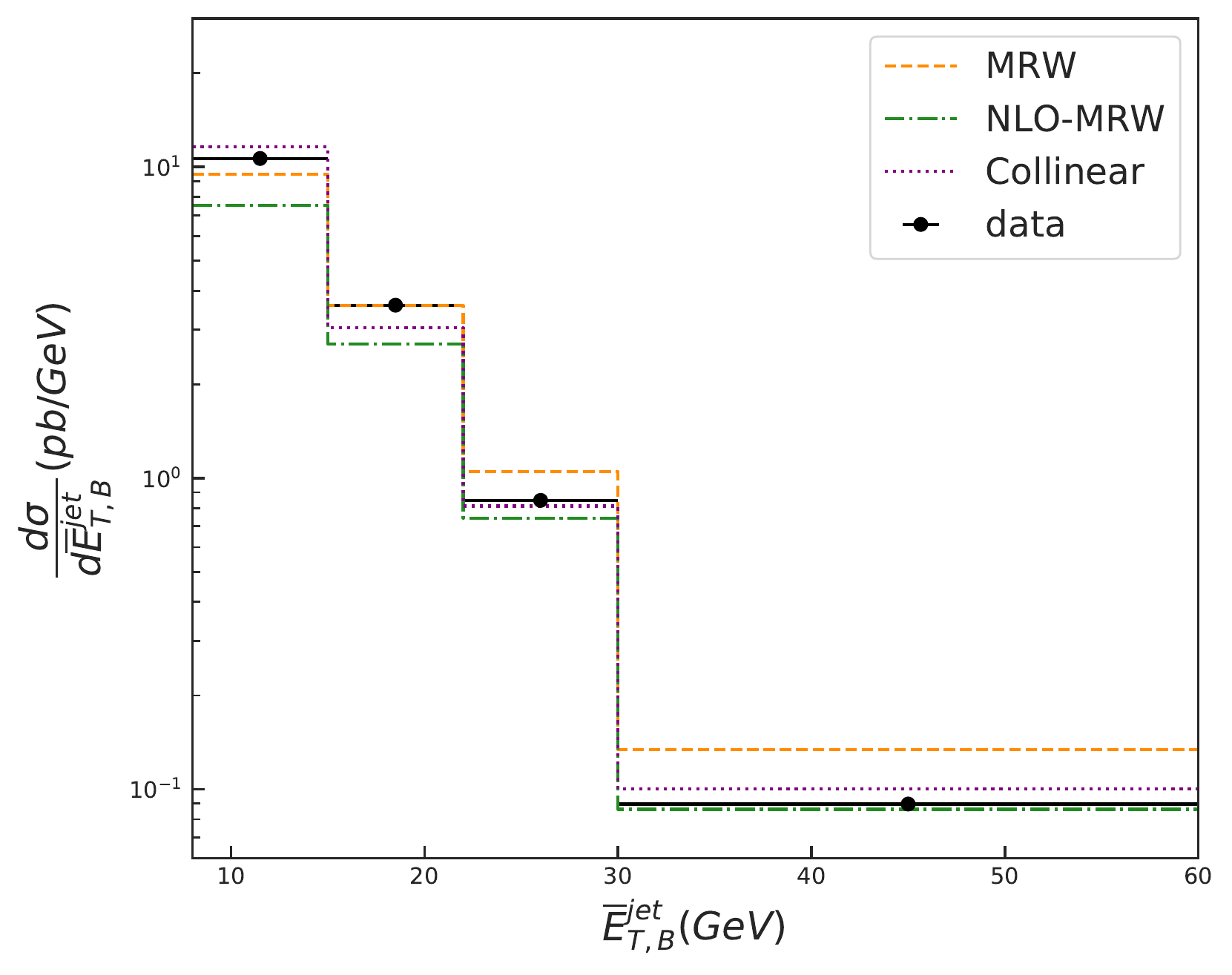}
		\caption
			{The comparison of   LO-MRW, NLO-MRW UPDFs $k_t$, and collinear factorizations  ep inclusive dijet  differential cross section with  those of  ZEUS collaboration data \cite{Zeus_2010}. The differential cross section is produced with the wrong (corrected) lab to the Breit transformation, the left panel (right panel), in the \textsc{KaTie}.
			} 
		\label{fig:1}
	\end{figure}

In the left panel of figure \ref{fig:1},  the differential  cross section with respect to the mean transverse jet energy in the Breit frame is obtained and compared   to the experimental data. As it can be seen in this figure, the result of the $k_t$-factorization with the LO-MRW and NLO-MRW UPDFs are strangely off the data. Because, in the previous works, It was shown  that the MRW UPDFs can produce the corresponding data reasonably \cite{Mod8,Mod9,Mod13,Mod14,Mod15,Mod16,Lipatov_photon,Baranov_Drell,lipatov_jet},  so to be sure what makes our calculation  to get  worse, we evaluated the differential cross section   within the collinear factorization framework, using the MMHT2014nlo68cl PDFs. The prediction of the collinear factorization, as depicted in the left panel of figure \ref{fig:1},   convinces us that there should be something wrong with this parton level event generator \cite{KATIE}. 
	
\begin{figure}
			\centering
		\includegraphics[width=16cm, height=14cm]{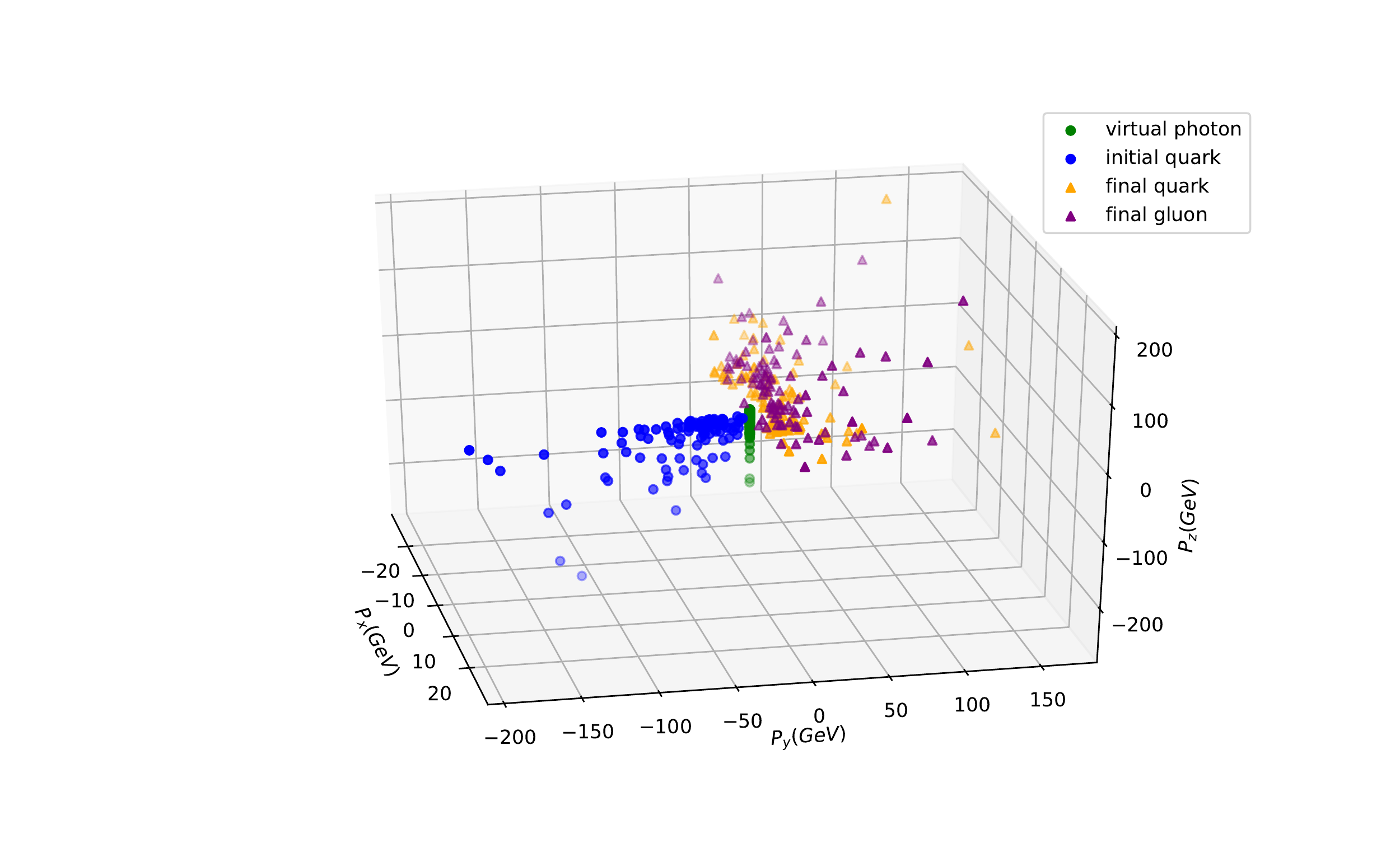}
		\caption
		{
			The three dimensional plot, with respect to the momenta $p_x$, $p_y$, and $p_z$, of $100$ events in the collinear factorization framework, for the sub-process $\gamma^\ast + q \to q + g$, generated by the actual \textsc{KaTie} parton level generator, before the Breit frame correction is made. 
		} 
		\label{fig:2}
	\end{figure}

In order to diagnose the reason for such results, we decided to visualize the three dimensional plot with respect to the momenta $p_x$, $p_y$, and $p_z$ of $100$ events generated by \textsc{KaTie}  in the Briet frame,  see the figure \ref{fig:2}. In this figure, the three dimensional momenta of   initial quark, virtual photon, final quark   and gluon of the sub-process, i.e., $\gamma^\ast + q \to q + g$, in the Breit frame within the collinear factorization are shown. In the collinear factorization framework, the virtual photon should collide head to head along the $z$ direction with the initial parton, according to     the Breit frame transformation, which does not happen as expected. According to the definition, in the Breit frame, the virtual photon has no energy component and only has momentum along the $-z$ direction, $q = (0, 0,0, -Q)$, and collinearly scatter head to head with quark  in the $+z$ direction \cite{Zeus_2010} which has momentum $xP$ ($P$ is the proton momentum). Therefore, it  is obvious that there is  problem    due to the wrong choice of lab to Breit frame transformation and it should be fixed \cite{Katie1}.
	
\begin{figure}
	\centering
	\includegraphics[width=16cm, height=14cm]{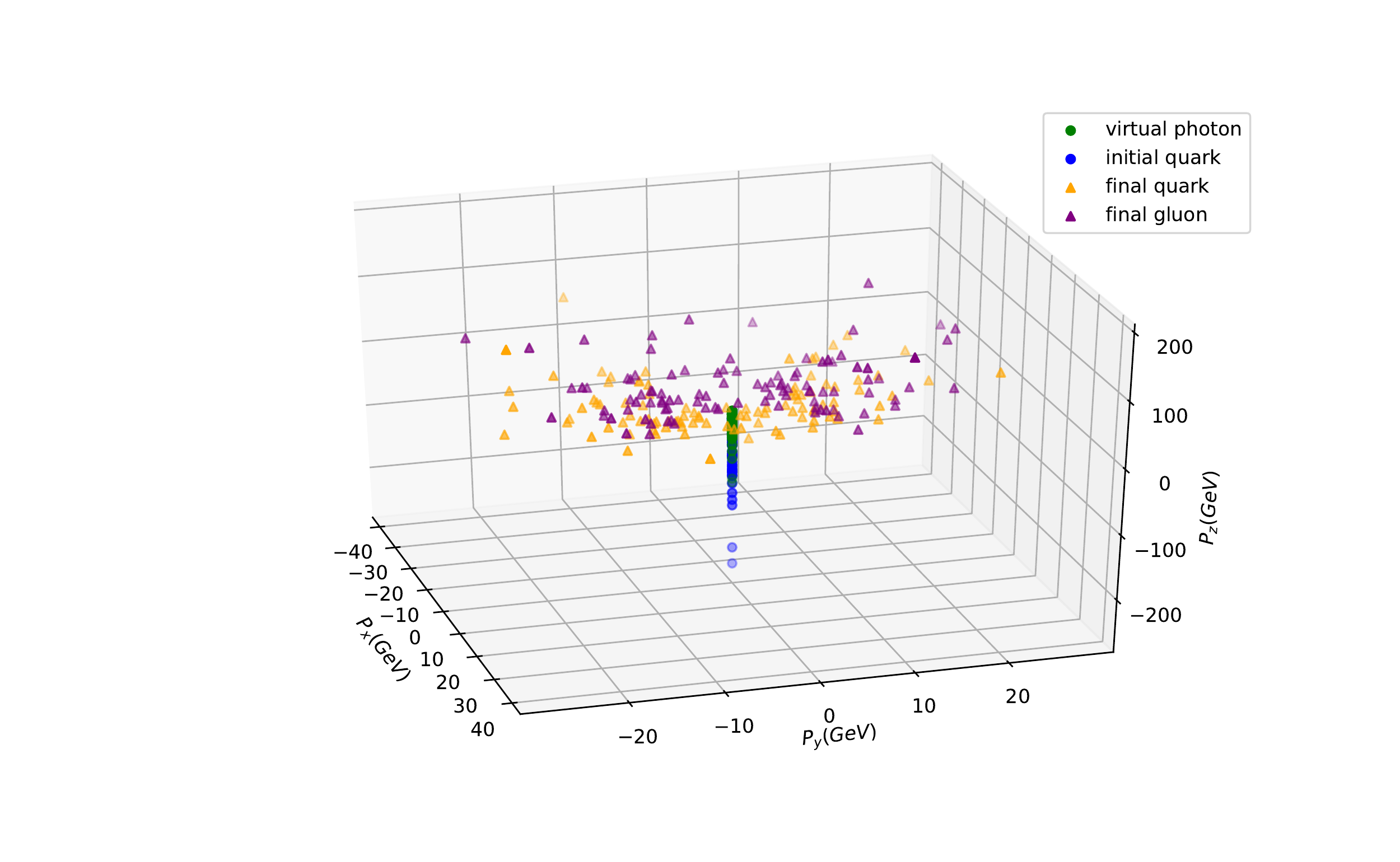}
	\caption
	{
		The three dimensional plot, with respect to the momenta $p_x$, $p_y$, and $p_z$, of $100$ events in the collinear factorization framework for the sub-process $\gamma^\ast + q \to q + g$  generated by the \textsc{KaTie} parton level generator   after correcting the lab to Breit transformation.
	} 
	\label{fig:3}
\end{figure}

This lab to Breit transformation can be simply fixed for example   according to  the reference \cite{Devenish:2004pb} (note that the minimum  requirement is ${\bf q}+2x{\bf P}=0$). If we make such corrections in the implementation of the \textsc{KaTie}, one can see that events are now behaving according to our expectation, see the figure \ref{fig:3}. Now, the virtual photon has a head to head collision with the initial quark along the $z$ direction. 
	
\begin{figure}
	\centering
	\includegraphics[width=8cm, height=8cm]{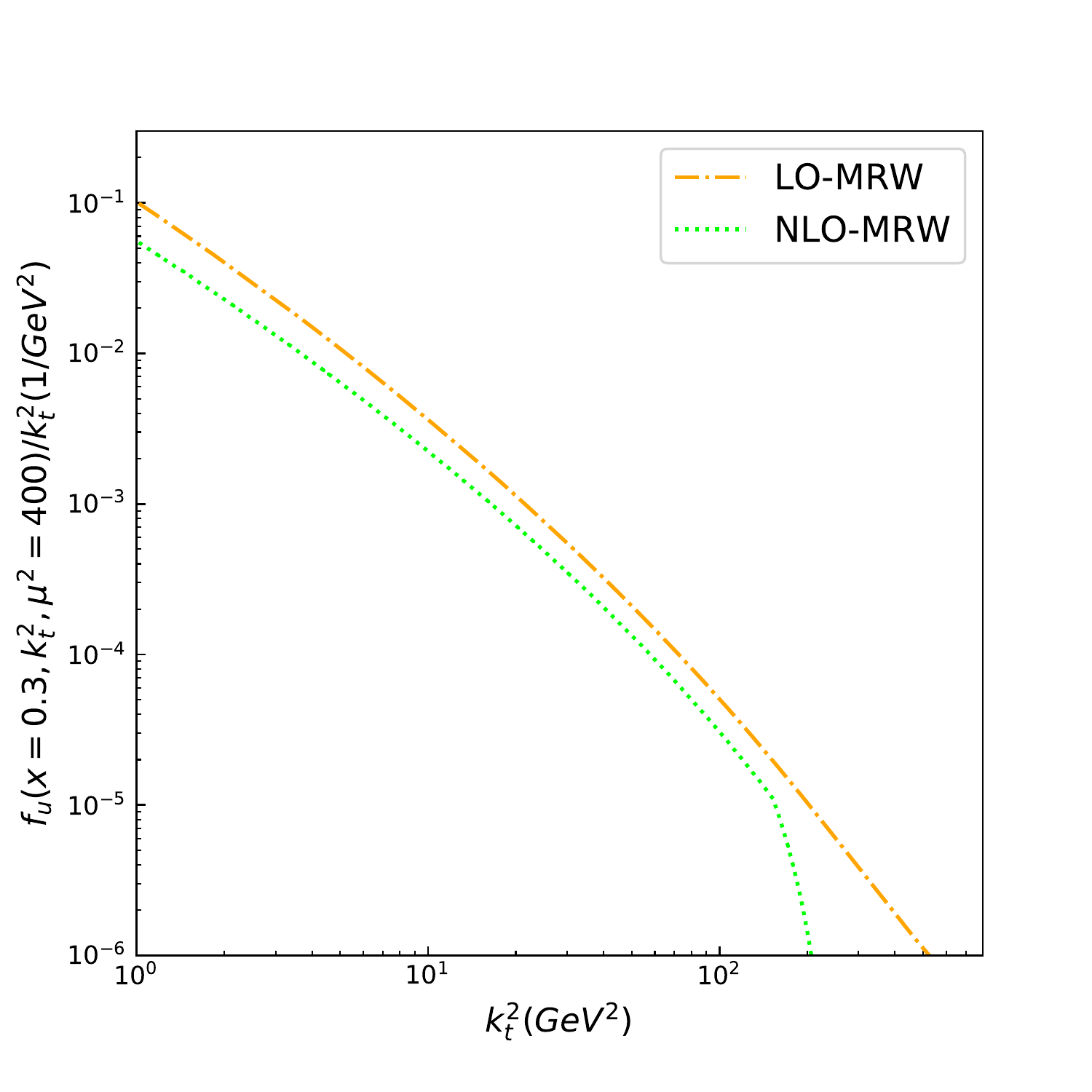}
	\includegraphics[width=8cm, height=8cm]{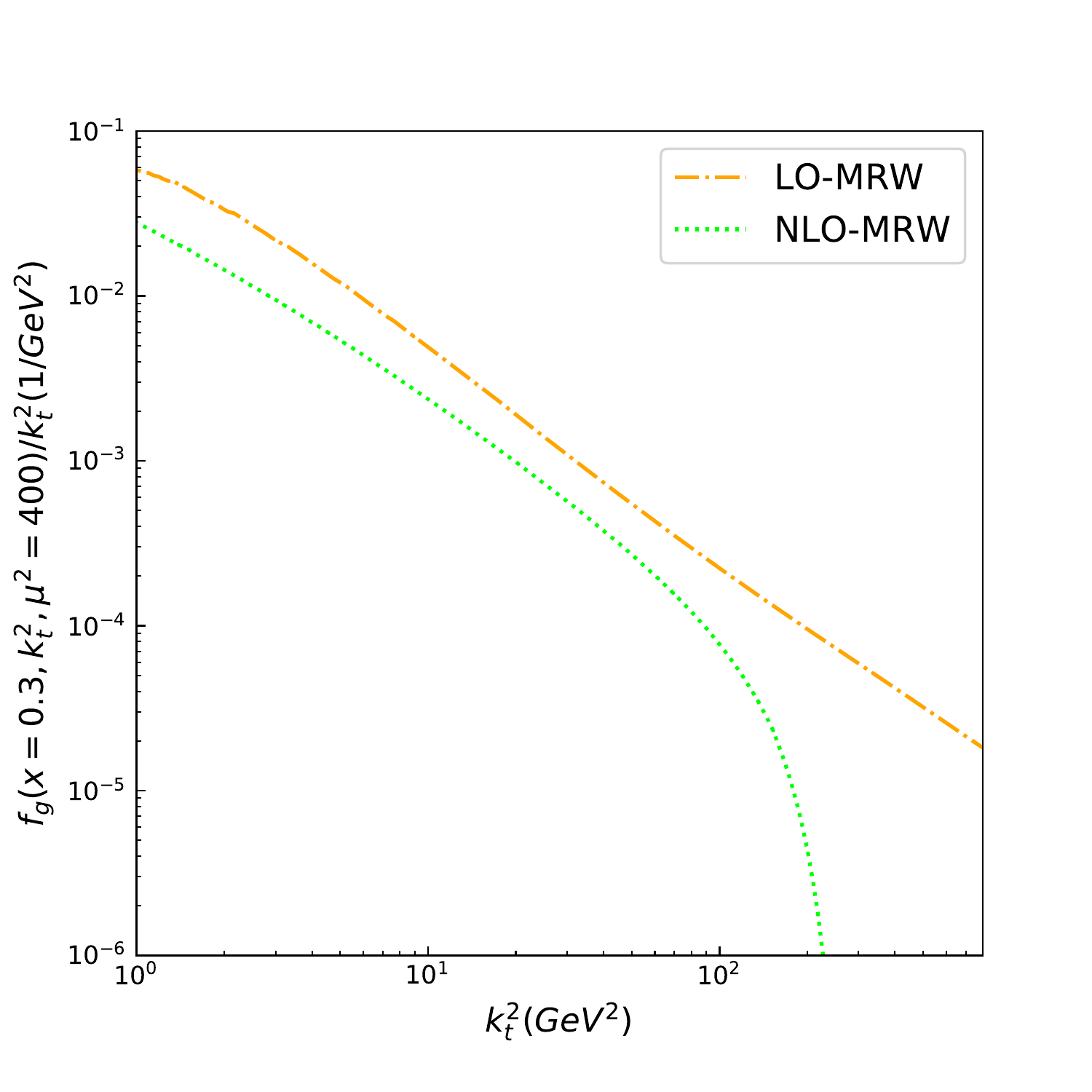}
	\includegraphics[width=8cm, height=8cm]{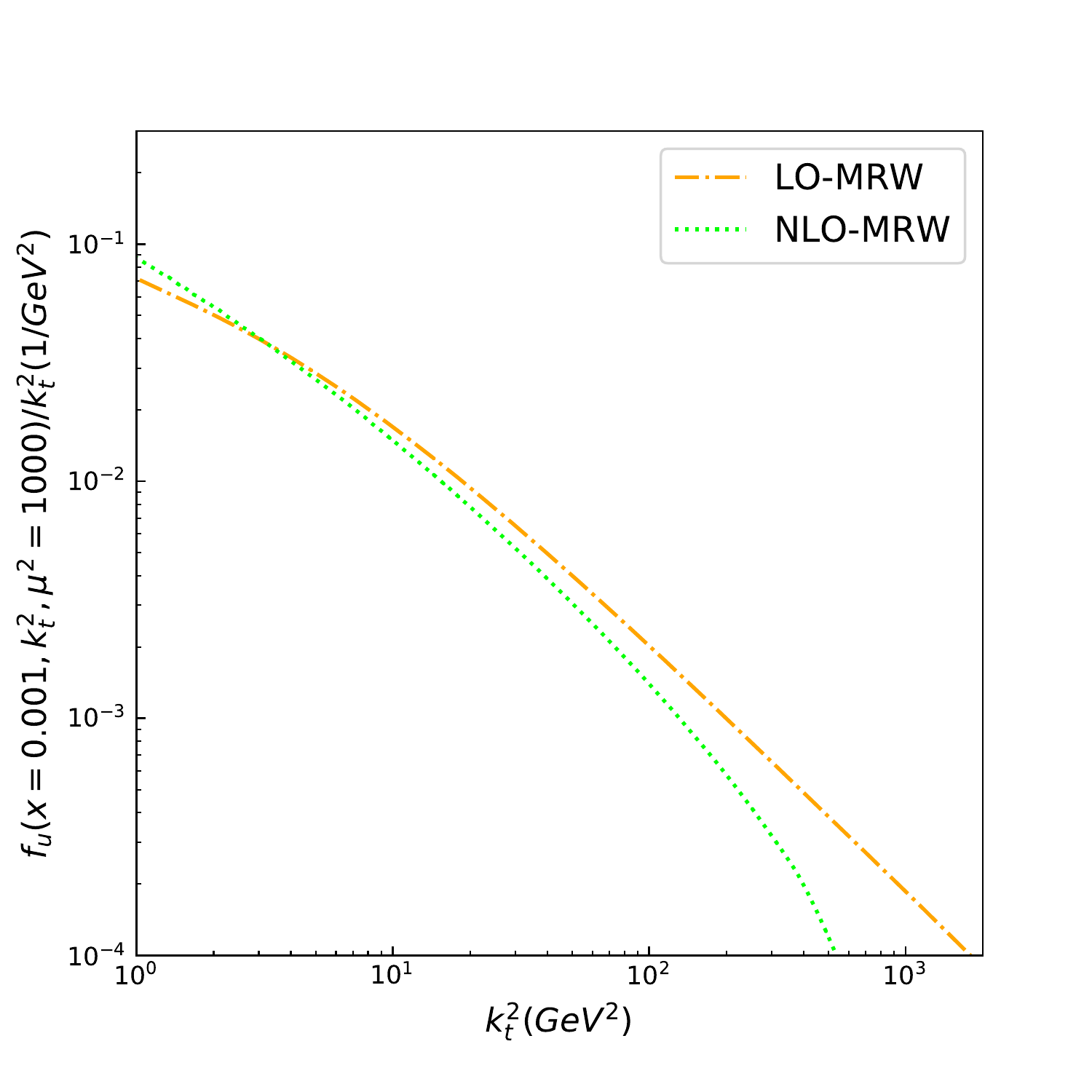}
		\includegraphics[width=8cm, height=8cm]{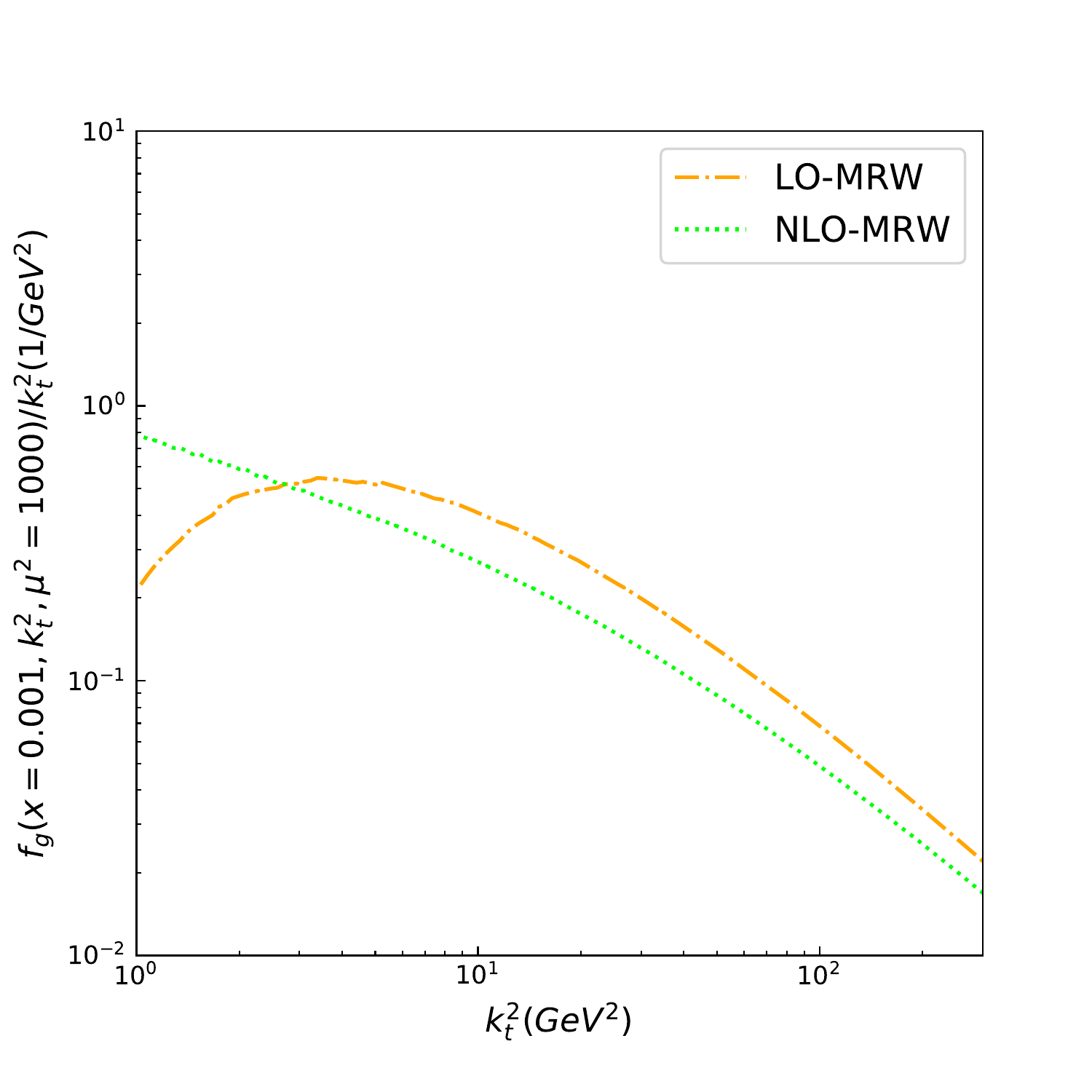}
	\caption
	{
		The left-upper (right-upper) panel shows the comparison of  up quark (gluon) LO-MRW and NLO-MRW UPDFs/$k_t^2$ at $x=0.3$ and $\mu^2=400$. The left-lower (right-lower) panel shows the comparison of  up quark (gluon) LO-MRW and NLO-MRW UPDFs/$k_t^2$ at $x=0.001$ and $\mu^2=1000$.
	} 
	\label{fig:4}
\end{figure}

Now we are in a position  to perform our calculation once again with the LO-MRW and NLO-MRW UPDFs, in addition to the collinear factorization framework witth the PDFs and compare them to the experimental data, see the right panel of figure \ref{fig:1}. Here, one can find that the collinear factorization can predict the data well, while the results of   LO-MRW UPDFs in the large mean transverse energy of final state jets overestimate the data, which is mostly due to the large UPDFs at large parton transverse momenta, see the reference \cite{PB_dynamical_res} for detail. Additionally,  the NLO-MRW UPDFs prediction, underestimate the data due to this fact that the choice of $k^2$ as a scale in this limit of energy, leads to the much larger scale than $k_t^2$ in the LO-MRW formalism. As a result of this fact and also the additional cutoff $\Theta(\mu^2-k^2)$ in this method, one could expect worse estimation of cross section for the NLO-MRW UPDFs    at this range of energy, see the top panels of figure \ref{fig:4} to gain an insight about these two different UPDFs for the up quark and the gluon at $x=0.3$ and $\mu^2=400$. However one should note that the $k_t$-factorization works much more better at small $x\le 0.01$ and very high center of mass energy $\ge 1 \ TeV$  which could be very useful for saving computer time \cite{Mod16}  with respect to NNLO collinear calculation in  future LHeC \cite{LHeC}.

\subsection{KATIE results for the ATLAS p-p Drell-Yan electron-pair production data} 
In this section we investigate the normalized differential cross section with respect to the transverse momentum of the Drell-Yan electron-pair production in the $k_t$-factorization framework in comparison with the ATLAS collaboration data \cite{Aad:2019wmn}. The data is related to the collision of proton-proton with center of mass energy $13\; TeV$. The electron (positron) transverse momenta are larger than $p_t^{e^-(e^+)} > 27 \; GeV$ with the invariant mass of the lepton pair $66\;GeV < M^{e^-e^+}< 116\;GeV$. Additionally, the absolute value of pseudo-rapidity of each electron(positron) are  $|\eta |< 2.47$, excluding $1.37< |\eta|<1.52$. For calculating the cross section we consider $q + \overline{q} \to e^- + e^+$ and $q + g \to e^- + e^+ + q$ sub-processes with the factorization and renormalization scale $\mu_R=\mu_F=\sqrt{M^{e^-e^+} + p_T^{e^-e^+}}$, with $n_f=5$.

\begin{figure}
	\centering
	\includegraphics[width=14cm, height=10cm]{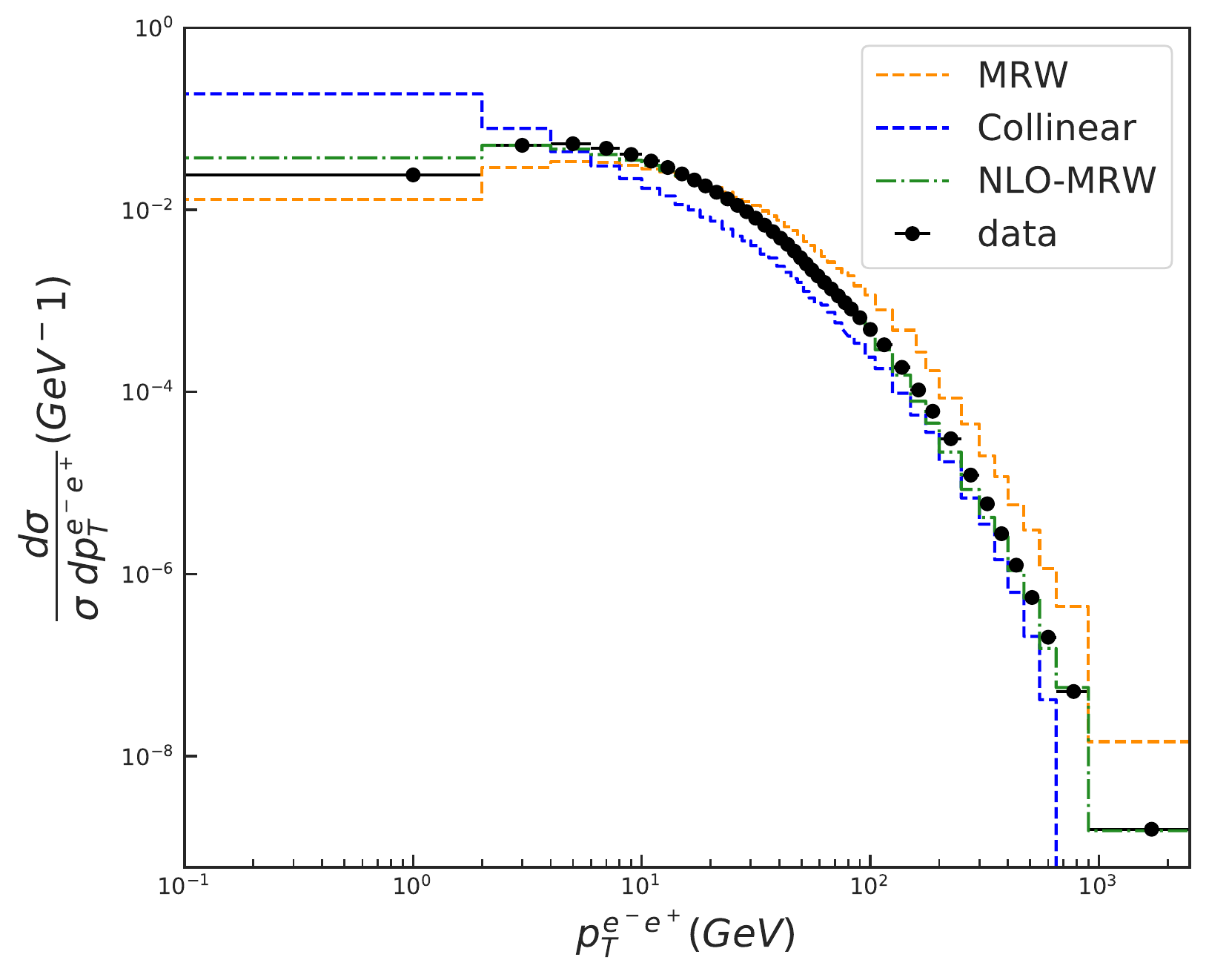}
	\caption
	{The comparison of   LO-MRW and NLO-MRW UPDFs $k_t$ factorization differential cross sections for the p-p Drell-Yan electron-pair production with the ATLAS collaboration data \cite{Aad:2019wmn} versus the transverse momentum of electron-positron, $p_T^{e^-e^+}$ . 
	} 
	\label{fig:5}
\end{figure}

In the figure \ref{fig:5}, we perform such comparison for the UPDFs of LO-MRW and NLO-MRW. It should be noted that the total cross section obtained with the LO-MRW and  NLO-MRW UPDFs are about $687.20\;pb$,   $683.79\;pb$, respectively, while the experimental cross section is measured to be about $738.3\;pb$. Here, we only mention the central values of the cross sections, because we did not calculate the uncertainty, see the table \ref{pred_lab_CS}. The NNLO and LO total cross sections are also tabulated.
As can be seen the total predicted cross sections using both the LO-MRW and NLO-MRW are close to the experimental measurement. However, one can find that in contrast to the ZEUS energy range, where the prediction of the NLO-MRW UPDFs undershoots the data, we obtain the results that satisfactory cover the data in most of the regions. While, regarding the figure \ref{fig:5}, the prediction of the LO-MRW still overshoots the data. For comparison, in the bottom panels of figure \ref{fig:4}, we also plot   $f_u(x=0.001, k_t^2, \mu^2=1000)/k_t^2$ and $f_g(x=0.001, k_t^2, \mu^2=1000)/k_t^2$   using the LO  and NLO-MRW UPDFs models. It should be noted that the suppression of NLO-MRW UPDFs, which happens at energy range of the ZEUS experiment, does not occur  for the LHC experiment. Because, at small $z$, the virtuality $k^2$ does not become much larger than the factorization scale, and hence the additional cutoff $\Theta(\mu^2-k^2)$ does not have much effect to make the result to undershoot the data.

To make a proper comparison, the collinear calculation with the \textsc{KaTie} for the above two processes is also plotted in the figure \ref{fig:5} which gives worst results with respect to the $k_t$-factorization formalisms, i.e., LO and NLO-MRW. One should note that the raise of the differential cross section at small $p_T^{e^-e^+}$ is due to incomplete 
  soft gluon resummation technique  which should be used to make QCD predictions at
low $p_T^{e^-e^+}$ finite \cite{LIP} and it does not happened in case of $k_t$-factorization calculations \cite{LIP} (in order to obtain finite results in case collinear calculation we imposed  $p_T^{e^-e^+}\ge 0.4 \ GeV$). The  application of NNLO collinear pQCD to the above differential cross section covers fully the data  \cite{Aad:2019wmn}. 

Finally, in contrast to our dijet calculations, one can find  that  the  application of $k_t$-factorization  in the \textsc{KaTie} parton level event generator, produce the  differential cross section consistent with the Drell-Yan experimental data. 
 \begin{table}[h]
 	\centering
 	\begin{tabular}[b]
 		{|c | c| c| c|}
 		\hline
 		  Model & $\sigma^{prediction} $  &    $\sigma^{experiment}$  &   $\sigma^{prediction} / \sigma^{experiment}$ \\
 		\hline
 		LO-MRW    & $683.79\;pb$            & $738.3\;pb$               &   $0.926$\\
 		\hline
 		NLO-MRW	  & $687.20\;pb$            & $738.3\;pb$               &   $0.930$\\
 		\hline  
 		Collinear-KATIE    & $586.03\;pb$            & $738.3\;pb$               &   $0.793$\\
 		\hline
 		NNLO  \cite{Aad:2019wmn}	  & $703.00\;pb$            & $738.3\;pb$               &   $0.952$\\
 		\hline  
 	\end{tabular}
 	\caption{The central   prediction  values of the LO-MRW,  NLO-MRW UPDFs $k_t$- and collinear factorizations of total ep inclusive  dijet cross-section in the fiducial volume in the electron decay channel, using  the \textsc{KaTie}, as well as the prediction/experiment of each model. The NNLO model is from the reference   \cite{Aad:2019wmn}}
 	\label{pred_lab_CS}
 \end{table}
 
\section{conclusion}
In this work, we investigated the \textsc{KaTie} parton level event generator for calculating the differential cross section of $ep$ inclusive dijet   and p-p Drell-Yan productions within the $k_t$- factorization framework in two energy ranges of the ZEUS and ATLAS collaboration data, respectively. In order to perform such calculation, we used the LO-MRW and NLO-MRW UPDFs with angular ordering constraint. For the $ep$ dijet  differential cross section,  it is noticed that the lab to Breit transformation is performed not correctly, in this parton level event generator, and should be fixed. After correcting this problem, we calculated the differential cross section and obtained acceptable results. We could obtain satisfactory results, when  the p-p Drell-Yan electron-pair production differential cross section in the above framework was calculated and compared with those of   ATLAS collaboration data. We understood that the NLO-MRW although underestimates    the ZEUS collaboration data, but can in fact predict the data of the ATLAS accurately. However, we observed that the LO-MRW overshoot the data in both the ZEUS and ATLAS collaboration data at large transverse momenta.  It was found, that the $k_t$-factorization is very useful tools for the small $x\le 0.01$ and very high center of mass energy $\ge 1 \ TeV$  and  could save much computer time \cite{Mod16}  with respect to NNLO collinear calculation,   especially in the future LHeC \cite{LHeC} and the present p-p LHC.

\newpage


\begin{thebibliography}{37}

	\expandafter\ifx\csname natexlab\endcsname\relax\def\natexlab#1{#1}\fi
	\expandafter\ifx\csname bibnamefont\endcsname\relax
	\def\bibnamefont#1{#1}\fi
	\expandafter\ifx\csname bibfnamefont\endcsname\relax
	\def\bibfnamefont#1{#1}\fi
	\expandafter\ifx\csname citenamefont\endcsname\relax
	\def\citenamefont#1{#1}\fi
	\expandafter\ifx\csname url\endcsname\relax
	\def\url#1{\texttt{#1}}\fi
	\expandafter\ifx\csname urlprefix\endcsname\relax\def\urlprefix{URL }\fi
	\providecommand{\bibinfo}[2]{#2}
	\providecommand{\eprint}[2][]{\url{#2}}
	
	\bibitem{dok}  Yu.L. Dokshitzer, V.A. Khoze, A.H. Mueller, S.I. Troyan,   Basics of Perturbative QCD, EDITIONS FRONTIERES, Singapore (1991).

\bibitem{Collins:2011zzd}J. Collins, Foundations of perturbative QCD, Cambridge University Press
 (2013). 
	
	\bibitem[{\citenamefont{van Hameren}(2018)}]{KATIE}
	\bibinfo{author}{\bibfnamefont{A.}~\bibnamefont{van Hameren}},
	\bibinfo{journal}{Comput. Phys. Commun.} \textbf{\bibinfo{volume}{224}},
	\bibinfo{pages}{371} (\bibinfo{year}{2018}).
	
	\bibitem[{\citenamefont{Alwall et~al.}(2007)}]{LHEF}
	\bibinfo{author}{\bibfnamefont{J.}~\bibnamefont{Alwall}} \bibnamefont{et~al.},
	\bibinfo{journal}{Comput. Phys. Commun.} \textbf{\bibinfo{volume}{176}},
	\bibinfo{pages}{300} (\bibinfo{year}{2007}).
	
	\bibitem{CASCADE3}S. Baranov et al, arXiv:2101.10221 (2021).
	
	\bibitem[{\citenamefont{Ciafaloni}(1988)}]{CCFM1}
	\bibinfo{author}{\bibfnamefont{M.}~\bibnamefont{Ciafaloni}},
	\bibinfo{journal}{Nuclear Physics B} \textbf{\bibinfo{volume}{296}},
	\bibinfo{pages}{49} (\bibinfo{year}{1988}).
	
	\bibitem[{\citenamefont{Catani et~al.}(1990{\natexlab{a}})\citenamefont{Catani,
			Fiorani, and Marchesini}}]{CCFM2}
	\bibinfo{author}{\bibfnamefont{S.}~\bibnamefont{Catani}},
	\bibinfo{author}{\bibfnamefont{F.}~\bibnamefont{Fiorani}}, \bibnamefont{and}
	\bibinfo{author}{\bibfnamefont{G.}~\bibnamefont{Marchesini}},
	\bibinfo{journal}{Nuclear Physics B} \textbf{\bibinfo{volume}{336}},
	\bibinfo{pages}{18} (\bibinfo{year}{1990}{\natexlab{a}}).
	
	\bibitem[{\citenamefont{Catani et~al.}(1990{\natexlab{b}})\citenamefont{Catani,
			Fiorani, and Marchesini}}]{CCFM3}
	\bibinfo{author}{\bibfnamefont{S.}~\bibnamefont{Catani}},
	\bibinfo{author}{\bibfnamefont{F.}~\bibnamefont{Fiorani}}, \bibnamefont{and}
	\bibinfo{author}{\bibfnamefont{G.}~\bibnamefont{Marchesini}},
	\bibinfo{journal}{Physics Letters B} \textbf{\bibinfo{volume}{234}},
	\bibinfo{pages}{339} (\bibinfo{year}{1990}{\natexlab{b}}).
	
	\bibitem[{\citenamefont{Marchesini}(1995)}]{CCFM4}
	\bibinfo{author}{\bibfnamefont{G.}~\bibnamefont{Marchesini}},
	\bibinfo{journal}{Nuclear Physics B} \textbf{\bibinfo{volume}{445}},
	\bibinfo{pages}{49} (\bibinfo{year}{1995}).
	
	\bibitem[{\citenamefont{Hautmann et~al.}(2018)\citenamefont{Hautmann, Jung,
			Lelek, Radescu, and Žlebčík}}]{PB1}
	\bibinfo{author}{\bibfnamefont{F.}~\bibnamefont{Hautmann}},
	\bibinfo{author}{\bibfnamefont{H.}~\bibnamefont{Jung}},
	\bibinfo{author}{\bibfnamefont{A.}~\bibnamefont{Lelek}},
	\bibinfo{author}{\bibfnamefont{V.}~\bibnamefont{Radescu}}, \bibnamefont{and}
	\bibinfo{author}{\bibfnamefont{R.}~\bibnamefont{Žlebčík}},
	\bibinfo{journal}{JHEP} \textbf{\bibinfo{volume}{2018}}, \bibinfo{pages}{70}
	(\bibinfo{year}{2018}).
	
	\bibitem[{\citenamefont{Hautmann et~al.}(2017)\citenamefont{Hautmann, Jung,
			Lelek, Radescu, and Žlebčík}}]{PB2}
	\bibinfo{author}{\bibfnamefont{F.}~\bibnamefont{Hautmann}},
	\bibinfo{author}{\bibfnamefont{H.}~\bibnamefont{Jung}},
	\bibinfo{author}{\bibfnamefont{A.}~\bibnamefont{Lelek}},
	\bibinfo{author}{\bibfnamefont{V.}~\bibnamefont{Radescu}}, \bibnamefont{and}
	\bibinfo{author}{\bibfnamefont{R.}~\bibnamefont{Žlebčík}},
	\bibinfo{journal}{Phys. Lett. B} \textbf{\bibinfo{volume}{772}},
	\bibinfo{pages}{446 } (\bibinfo{year}{2017}).
	
	\bibitem[{\citenamefont{Dokshitzer}(1977)}]{DokshitzerDeepInelastic}
	\bibinfo{author}{\bibfnamefont{Y.~L.} \bibnamefont{Dokshitzer}},
	\bibinfo{journal}{Sov. Phys. JETP} \textbf{\bibinfo{volume}{46}},
	\bibinfo{pages}{641} (\bibinfo{year}{1977}).
	
	\bibitem[{\citenamefont{Gribov and Lipatov}(1972)}]{gribovDeepInelastic}
	\bibinfo{author}{\bibfnamefont{V.~N.} \bibnamefont{Gribov}} \bibnamefont{and}
	\bibinfo{author}{\bibfnamefont{L.~N.} \bibnamefont{Lipatov}},
	\bibinfo{journal}{Yad. Fiz.} \textbf{\bibinfo{volume}{15}},
	\bibinfo{pages}{781} (\bibinfo{year}{1972}).
	
	\bibitem[{\citenamefont{Altarelli and Parisi}(1977)}]{altarelli}
	\bibinfo{author}{\bibfnamefont{G.}~\bibnamefont{Altarelli}} \bibnamefont{and}
	\bibinfo{author}{\bibfnamefont{G.}~\bibnamefont{Parisi}},
	\bibinfo{journal}{Nucl. Phys. B} \textbf{\bibinfo{volume}{126}},
	\bibinfo{pages}{298} (\bibinfo{year}{1977}).
	
	\bibitem[{\citenamefont{Hautmann et~al.}(2019)\citenamefont{Hautmann,
			Keersmaekers, Lelek, and van Kampen}}]{PB_dynamical_res}
	\bibinfo{author}{\bibfnamefont{F.}~\bibnamefont{Hautmann}},
	\bibinfo{author}{\bibfnamefont{L.}~\bibnamefont{Keersmaekers}},
	\bibinfo{author}{\bibfnamefont{A.}~\bibnamefont{Lelek}}, \bibnamefont{and}
	\bibinfo{author}{\bibfnamefont{A.}~\bibnamefont{van Kampen}},
	\bibinfo{journal}{Nucl. Phys. B} \textbf{\bibinfo{volume}{949}},
	\bibinfo{pages}{114795} (\bibinfo{year}{2019}).
	
	\bibitem[{\citenamefont{Bermudez~Martinez et~al.}(2019)}]{PB_application}
	\bibinfo{author}{\bibfnamefont{A.}~\bibnamefont{Bermudez~Martinez}}
	\bibnamefont{et~al.}, \bibinfo{journal}{Phys. Rev. D}
	\textbf{\bibinfo{volume}{100}}, \bibinfo{pages}{074027}
	(\bibinfo{year}{2019}).
	
	\bibitem[{\citenamefont{Martin et~al.}(2010)\citenamefont{Martin, Ryskin, and
			Watt}}]{MRW}
	\bibinfo{author}{\bibfnamefont{A.~D.} \bibnamefont{Martin}},
	\bibinfo{author}{\bibfnamefont{M.~G.} \bibnamefont{Ryskin}},
	\bibnamefont{and} \bibinfo{author}{\bibfnamefont{G.}~\bibnamefont{Watt}},
	\bibinfo{journal}{Eur. Phys. J. C} \textbf{\bibinfo{volume}{66}},
	\bibinfo{pages}{163} (\bibinfo{year}{2010}).
	
	\bibitem{MRW1} G. Watt, A.D. Martin, M.G. Ryskin, Eur. Phys. J. C, {\bf 31}, 73 (2003).
	
	\bibitem[{\citenamefont{Modarres and Hosseinkhani}(2009)}]{Mod1}
	\bibinfo{author}{\bibfnamefont{M.}~\bibnamefont{Modarres}} \bibnamefont{and}
	\bibinfo{author}{\bibfnamefont{H.}~\bibnamefont{Hosseinkhani}},
	\bibinfo{journal}{Nucl.Phys.A} \textbf{\bibinfo{volume}{815}},
	\bibinfo{pages}{40} (\bibinfo{year}{2009}).
	
	\bibitem[{\citenamefont{Modarres and Hosseinkhani}(2010)}]{Mod2}
	\bibinfo{author}{\bibfnamefont{M.}~\bibnamefont{Modarres}} \bibnamefont{and}
	\bibinfo{author}{\bibfnamefont{H.}~\bibnamefont{Hosseinkhani}},
	\bibinfo{journal}{Few-Body Syst.} \textbf{\bibinfo{volume}{47}},
	\bibinfo{pages}{237} (\bibinfo{year}{2010}).
	
	\bibitem[{\citenamefont{Hosseinkhani and Modarres}(2011)}]{Mod3}
	\bibinfo{author}{\bibfnamefont{H.}~\bibnamefont{Hosseinkhani}}
	\bibnamefont{and} \bibinfo{author}{\bibfnamefont{M.}~\bibnamefont{Modarres}},
	\bibinfo{journal}{Phys.Lett.B} \textbf{\bibinfo{volume}{694}},
	\bibinfo{pages}{355} (\bibinfo{year}{2011}).
	
	\bibitem[{\citenamefont{Hosseinkhani and Modarres}(2012)}]{Mod4}
	\bibinfo{author}{\bibfnamefont{H.}~\bibnamefont{Hosseinkhani}}
	\bibnamefont{and} \bibinfo{author}{\bibfnamefont{M.}~\bibnamefont{Modarres}},
	\bibinfo{journal}{Phys.Lett.B} \textbf{\bibinfo{volume}{708}},
	\bibinfo{pages}{75} (\bibinfo{year}{2012}).
	
	\bibitem[{\citenamefont{Modarres et~al.}(2013)\citenamefont{Modarres,
			Hosseinkhani, and Olanj}}]{Mod5}
	\bibinfo{author}{\bibfnamefont{M.}~\bibnamefont{Modarres}},
	\bibinfo{author}{\bibfnamefont{H.}~\bibnamefont{Hosseinkhani}},
	\bibnamefont{and} \bibinfo{author}{\bibfnamefont{N.}~\bibnamefont{Olanj}},
	\bibinfo{journal}{Nucl.Phys.A} \textbf{\bibinfo{volume}{902}},
	\bibinfo{pages}{21} (\bibinfo{year}{2013}).
	
	\bibitem[{\citenamefont{Modarres et~al.}(2014)\citenamefont{Modarres,
			Hosseinkhani, and Olanj}}]{Mod6}
	\bibinfo{author}{\bibfnamefont{M.}~\bibnamefont{Modarres}},
	\bibinfo{author}{\bibfnamefont{H.}~\bibnamefont{Hosseinkhani}},
	\bibnamefont{and} \bibinfo{author}{\bibfnamefont{N.}~\bibnamefont{Olanj}},
	\bibinfo{journal}{Phys.Rev.D} \textbf{\bibinfo{volume}{89}},
	\bibinfo{pages}{034015} (\bibinfo{year}{2014}).
	
	\bibitem[{\citenamefont{Modarres
			et~al.}(2017{\natexlab{a}})\citenamefont{Modarres, Hosseinkhani, and
			Olanj}}]{Mod7}
	\bibinfo{author}{\bibfnamefont{M.}~\bibnamefont{Modarres}},
	\bibinfo{author}{\bibfnamefont{H.}~\bibnamefont{Hosseinkhani}},
	\bibnamefont{and} \bibinfo{author}{\bibfnamefont{N.}~\bibnamefont{Olanj}},
	\bibinfo{journal}{Int.J.Mod.Phys.A} \textbf{\bibinfo{volume}{32}},
	\bibinfo{pages}{1750121} (\bibinfo{year}{2017}{\natexlab{a}}).
	
	\bibitem[{\citenamefont{Modarres et~al.}(2015)\citenamefont{Modarres,
			Hosseinkhani, Olanj, and Masouminia}}]{Mod8}
	\bibinfo{author}{\bibfnamefont{M.}~\bibnamefont{Modarres}},
	\bibinfo{author}{\bibfnamefont{H.}~\bibnamefont{Hosseinkhani}},
	\bibinfo{author}{\bibfnamefont{N.}~\bibnamefont{Olanj}}, \bibnamefont{and}
	\bibinfo{author}{\bibfnamefont{M.}~\bibnamefont{Masouminia}},
	\bibinfo{journal}{Eur.Phys.J.C} \textbf{\bibinfo{volume}{75}},
	\bibinfo{pages}{556} (\bibinfo{year}{2015}).
	
	\bibitem[{\citenamefont{Modarres et~al.}(2016)\citenamefont{Modarres,
			Masouminia, Hosseinkhani, and Olanj}}]{Mod9}
	\bibinfo{author}{\bibfnamefont{M.}~\bibnamefont{Modarres}},
	\bibinfo{author}{\bibfnamefont{M.}~\bibnamefont{Masouminia}},
	\bibinfo{author}{\bibfnamefont{H.}~\bibnamefont{Hosseinkhani}},
	\bibnamefont{and} \bibinfo{author}{\bibfnamefont{N.}~\bibnamefont{Olanj}},
	\bibinfo{journal}{Nucl.Phys.A} \textbf{\bibinfo{volume}{945}},
	\bibinfo{pages}{168} (\bibinfo{year}{2016}).
	
	\bibitem[{\citenamefont{Modarres
			et~al.}(2017{\natexlab{b}})\citenamefont{Modarres, Masouminia, Aminzadeh~Nik,
			Hosseinkhani, and Olanj}}]{Mod13}
	\bibinfo{author}{\bibfnamefont{M.}~\bibnamefont{Modarres}},
	\bibinfo{author}{\bibfnamefont{M.}~\bibnamefont{Masouminia}},
	\bibinfo{author}{\bibfnamefont{R.}~\bibnamefont{Aminzadeh~Nik}},
	\bibinfo{author}{\bibfnamefont{H.}~\bibnamefont{Hosseinkhani}},
	\bibnamefont{and} \bibinfo{author}{\bibfnamefont{N.}~\bibnamefont{Olanj}},
	\bibinfo{journal}{Nucl.Phys.B} \textbf{\bibinfo{volume}{922}},
	\bibinfo{pages}{94} (\bibinfo{year}{2017}{\natexlab{b}}).
	
	\bibitem[{\citenamefont{Modarres et~al.}(2018)\citenamefont{Modarres,
			Masouminia, Aminzadeh~Nik, Hosseinkhani, and Olanj}}]{Mod14}
	\bibinfo{author}{\bibfnamefont{M.}~\bibnamefont{Modarres}},
	\bibinfo{author}{\bibfnamefont{M.}~\bibnamefont{Masouminia}},
	\bibinfo{author}{\bibfnamefont{R.}~\bibnamefont{Aminzadeh~Nik}},
	\bibinfo{author}{\bibfnamefont{H.}~\bibnamefont{Hosseinkhani}},
	\bibnamefont{and} \bibinfo{author}{\bibfnamefont{N.}~\bibnamefont{Olanj}},
	\bibinfo{journal}{Nucl. Phys. B} \textbf{\bibinfo{volume}{926}},
	\bibinfo{pages}{406} (\bibinfo{year}{2018}).
	
	\bibitem[{\citenamefont{Aminzadeh~Nik et~al.}(2018)\citenamefont{Aminzadeh~Nik,
			Modarres, and Masouminia}}]{Mod15}
	\bibinfo{author}{\bibfnamefont{R.}~\bibnamefont{Aminzadeh~Nik}},
	\bibinfo{author}{\bibfnamefont{M.}~\bibnamefont{Modarres}}, \bibnamefont{and}
	\bibinfo{author}{\bibfnamefont{M.~R.} \bibnamefont{Masouminia}},
	\bibinfo{journal}{Phys. Rev. D} \textbf{\bibinfo{volume}{97}},
	\bibinfo{pages}{096012} (\bibinfo{year}{2018}).
	
	\bibitem[{\citenamefont{Modarres et~al.}(2019)\citenamefont{Modarres,
			Aminzadeh~Nik, Kord~Valeshbadi, Hosseinkhani, and Olanj}}]{Mod16}
	\bibinfo{author}{\bibfnamefont{M.}~\bibnamefont{Modarres}},
	\bibinfo{author}{\bibfnamefont{R.}~\bibnamefont{Aminzadeh~Nik}},
	\bibinfo{author}{\bibfnamefont{R.}~\bibnamefont{Kord~Valeshbadi}},
	\bibinfo{author}{\bibfnamefont{H.}~\bibnamefont{Hosseinkhani}},
	\bibnamefont{and} \bibinfo{author}{\bibfnamefont{N.}~\bibnamefont{Olanj}},
	\bibinfo{journal}{J. Phys. G} \textbf{\bibinfo{volume}{46}},
	\bibinfo{pages}{105005} (\bibinfo{year}{2019}).
	
	\bibitem[{\citenamefont{Lipatov and Malyshev}(2016)}]{Lipatov_photon}
	\bibinfo{author}{\bibfnamefont{A.~V.} \bibnamefont{Lipatov}} \bibnamefont{and}
	\bibinfo{author}{\bibfnamefont{M.~A.} \bibnamefont{Malyshev}},
	\bibinfo{journal}{Phys. Rev. D} \textbf{\bibinfo{volume}{94}},
	\bibinfo{pages}{034020} (\bibinfo{year}{2016}).
	
	\bibitem[{\citenamefont{Baranov et~al.}(2014)\citenamefont{Baranov, Lipatov,
			and Zotov}}]{Baranov_Drell}
	\bibinfo{author}{\bibfnamefont{S.~P.} \bibnamefont{Baranov}},
	\bibinfo{author}{\bibfnamefont{A.~V.} \bibnamefont{Lipatov}},
	\bibnamefont{and} \bibinfo{author}{\bibfnamefont{N.~P.} \bibnamefont{Zotov}},
	\bibinfo{journal}{Phys. Rev. D} \textbf{\bibinfo{volume}{89}},
	\bibinfo{pages}{094025} (\bibinfo{year}{2014}).
	
	\bibitem[{\citenamefont{Lipatov and Zotov}(2014)}]{lipatov_jet}
	\bibinfo{author}{\bibfnamefont{A.~V.} \bibnamefont{Lipatov}} \bibnamefont{and}
	\bibinfo{author}{\bibfnamefont{N.~P.} \bibnamefont{Zotov}},
	\bibinfo{journal}{Phys. Rev. D} \textbf{\bibinfo{volume}{90}},
	\bibinfo{pages}{094005} (\bibinfo{year}{2014}).
	
	
\bibitem[{\citenamefont{van Hameren}(2019)}]{vanHameren:2019wzx}
\bibinfo{author}{\bibfnamefont{A.}~\bibnamefont{van Hameren}},
\bibinfo{journal}{PoS} \textbf{\bibinfo{volume}{DIS2019}},
\bibinfo{pages}{139} (\bibinfo{year}{2019}).

\bibitem{inc-dijet-mod} R. Kord Valeshabadi, M. Modarres, S. Rezaie, R. Aminzadeh Nik, J.Phys.G (2021) submitted for publication.

\bibitem[{\citenamefont{Abramowicz et~al.}(2010)\citenamefont{Abramowicz, Abt,
			Adamczyk et~al.}}]{Zeus_2010}
	\bibinfo{author}{\bibfnamefont{H.}~\bibnamefont{Abramowicz}},
	\bibinfo{author}{\bibfnamefont{I.}~\bibnamefont{Abt}},
	\bibinfo{author}{\bibnamefont{Adamczyk}}, \bibnamefont{et~al.},
	\bibinfo{journal}{Eur. Phys. J. C} \textbf{\bibinfo{volume}{70}},
	\bibinfo{pages}{965} (\bibinfo{year}{2010}).
	
	\bibitem[{\citenamefont{Aad et~al.}(2020)}]{Aad:2019wmn}
	\bibinfo{author}{\bibfnamefont{G.}~\bibnamefont{Aad}} \bibnamefont{et~al.}
	(\bibinfo{collaboration}{ATLAS}), \bibinfo{journal}{Eur. Phys. J. C}
	\textbf{\bibinfo{volume}{80}}, \bibinfo{pages}{616} (\bibinfo{year}{2020}).
	
	\bibitem[{\citenamefont{Hautmann et~al.}(2014)\citenamefont{Hautmann, Jung
			et~al.}}]{TMDLIB}
	\bibinfo{author}{\bibfnamefont{F.}~\bibnamefont{Hautmann}},
	\bibinfo{author}{\bibnamefont{Jung}}, \bibnamefont{et~al.},
	\bibinfo{journal}{Eur. Phys. J. C} \textbf{\bibinfo{volume}{74}}
	(\bibinfo{year}{2014}).
	
	\bibitem[{\citenamefont{Baranov
			et~al.}(2021{\natexlab{b}})\citenamefont{Baranov, Martinez, Banos, Guzman,
			Hautmann, Jung, Lelek, Lidrych, Lipatov, Malyshev
			et~al.}}]{baranov2021cascade3}
	\bibinfo{author}{\bibfnamefont{S.}~\bibnamefont{Baranov}},
	\bibinfo{author}{\bibfnamefont{A.~B.} \bibnamefont{Martinez}},
	\bibinfo{author}{\bibfnamefont{L.~I.~E.} \bibnamefont{Banos}},
	\bibinfo{author}{\bibfnamefont{F.}~\bibnamefont{Guzman}},
	\bibinfo{author}{\bibfnamefont{F.}~\bibnamefont{Hautmann}},
	\bibinfo{author}{\bibfnamefont{H.}~\bibnamefont{Jung}},
	\bibinfo{author}{\bibfnamefont{A.}~\bibnamefont{Lelek}},
	\bibinfo{author}{\bibfnamefont{J.}~\bibnamefont{Lidrych}},
	\bibinfo{author}{\bibfnamefont{A.}~\bibnamefont{Lipatov}},
	\bibinfo{author}{\bibfnamefont{M.}~\bibnamefont{Malyshev}},
	\bibnamefont{et~al.}, \emph{\bibinfo{title}{Cascade3 a monte carlo event
			generator based on tmds}} (\bibinfo{year}{2021}{\natexlab{b}}),
	\eprint{2101.10221}.
	
	
\bibitem[{\citenamefont{Harland-Lang et~al.}(2015)\citenamefont{Harland-Lang,
Martin, Motylinski, and Thorne}}]{harland-lang_uncertainties_2015}
\bibinfo{author}{\bibfnamefont{L.~A.} \bibnamefont{Harland-Lang}},
\bibinfo{author}{\bibfnamefont{A.~D.} \bibnamefont{Martin}},
\bibinfo{author}{\bibfnamefont{P.}~\bibnamefont{Motylinski}},
\bibnamefont{and} \bibinfo{author}{\bibfnamefont{R.~S.}
\bibnamefont{Thorne}}, \bibinfo{journal}{Eur. Phys. J. C}
\textbf{\bibinfo{volume}{75}}, \bibinfo{pages}{435} (\bibinfo{year}{2015}).

\bibitem[{\citenamefont{Harland-Lang et~al.}(2016)\citenamefont{Harland-Lang,
Martin, Motylinski, and Thorne}}]{harland-lang_charm_2016}
\bibinfo{author}{\bibfnamefont{L.~A.} \bibnamefont{Harland-Lang}},
\bibinfo{author}{\bibfnamefont{A.~D.} \bibnamefont{Martin}},
\bibinfo{author}{\bibfnamefont{P.}~\bibnamefont{Motylinski}},
\bibnamefont{and} \bibinfo{author}{\bibfnamefont{R.~S.}
\bibnamefont{Thorne}}, \bibinfo{journal}{Eur. Phys. J. C}
\textbf{\bibinfo{volume}{76}}, \bibinfo{pages}{10} (\bibinfo{year}{2016}).


\bibitem[{\citenamefont{Buckley et~al.}(2015)}]{LHAPDF6}
\bibinfo{author}{\bibfnamefont{A.}~\bibnamefont{Buckley}} \bibnamefont{et~al.},
\bibinfo{journal}{Eur. Phys. J. C} \textbf{\bibinfo{volume}{75}},
\bibinfo{pages}{132} (\bibinfo{year}{2015}).
	
	
	
	\bibitem[{\citenamefont{van Hameren}()}]{Katie1}
	\bibinfo{author}{\bibfnamefont{A.}~\bibnamefont{van Hameren}},
	\bibinfo{title}{private communication.}
	
	\bibitem[{\citenamefont{Devenish and Cooper-Sarkar}(2004)}]{Devenish:2004pb}
	\bibinfo{author}{\bibfnamefont{R.}~\bibnamefont{Devenish}} \bibnamefont{and}
	\bibinfo{author}{\bibfnamefont{A.}~\bibnamefont{Cooper-Sarkar}},
	\emph{\bibinfo{title}{{Deep inelastic scattering}}} (\bibinfo{year}{2004}).
	

\bibitem{LHeC}LHeC Study Group, J L Abelleira Fernandez et al, J.Phys.G:Nucl.Part.Phys., {\bf 39} 075001	(2012).

\bibitem{LIP} A.V. Lipatov, M.A.  Malyshev and N.P. Zotov,  JHEP, {\bf 2011} 117 (2011).
	
\end{thebibliography}
\end{document}